\pdfoutput=1

\documentclass[11pt]{article}

\usepackage[final]{acl}
\usepackage{times}
\usepackage{latexsym}

\usepackage[T1]{fontenc}
\usepackage{amsmath}
\usepackage{amsthm}
\usepackage[utf8]{inputenc}
\usepackage{microtype}
\usepackage{inconsolata}
\usepackage{graphicx}

\usepackage{amssymb}
\usepackage{enumitem}
\usepackage{euscript} 
\usepackage{graphicx}
\usepackage{colortbl}
\usepackage{multirow}
\usepackage{xcolor}
\usepackage{tcolorbox} 
\usepackage{booktabs}
\usepackage{subfig}
\usepackage{lipsum}
\usepackage{algorithm}
\usepackage{algorithmic}
\newtheorem{theorem}{Theorem}

\usepackage[normalem]{ulem}

\title{$K$-order Ranking Preference Optimization for Large Language Models}
\author{
    \textbf{Shihao Cai\textsuperscript{1}},
    \textbf{Chongming Gao\textsuperscript{1}} \footnotemark[2],
    \textbf{Yang Zhang\textsuperscript{2}},
    \textbf{Wentao Shi\textsuperscript{1}},
    \\
    \textbf{Jizhi Zhang\textsuperscript{1}} ,
    \textbf{Keqin Bao\textsuperscript{1}} ,
    \textbf{Qifan Wang\textsuperscript{3}} ,
    \textbf{Fuli Feng\textsuperscript{1}} \footnotemark[2]
    \\
    \\
    \textsuperscript{1} University of Science and Technology of China,
    \textsuperscript{2} National University of Singapore,
    \textsuperscript{3} Meta AI
    \\
    {
        caishihao@mail.ustc.edu.cn, \{chongming.gao,  zyang1580\}@gmail.com
    }
    \\{
        \{shiwentao123, baokq, cdzhangjizhi\}@mail.ustc.edu.cn 
    }
    \\{
        wqfcr@fb.com, fulifeng93@gmail.com
    }
}

\begin{document}
\maketitle
\newcommand\blfootnote[1]{%
\begingroup
\renewcommand\thefootnote{}\footnote{#1}%
\addtocounter{footnote}{-1}%
\endgroup
}

\blfootnote{\dag Corresponding authors.}
\begin{abstract}
To adapt large language models (LLMs) to ranking tasks, existing list-wise methods, represented by list-wise Direct Preference Optimization (DPO), focus on optimizing partial-order or full-order list ranking consistency for LLMs to enhance their ranking abilities.
However, we argue that optimizing top-K ranking consistency could be more appropriate for real-world applications. There are two main reasons: (1) users are typically concerned with only the top-K results, making top-K ranking more important, and (2) tail items often lack precise feedback, making top-K ranking more reliable. Based on this, we propose $\textit{\textbf{K}}$-order Ranking \textbf{P}reference \textbf{O}ptimization (KPO) by extending the DPO’s Plackett-Luce model to accommodate top-K rankings. Additionally, recognizing that the number of important items can vary across queries, we extend KPO to dynamically determine appropriate $K$ for different samples and introduce a curriculum learning strategy to boost training efficiency. Extensive experiments demonstrate the effectiveness of KPO, highlighting its high sample efficiency and robustness to noise. The code is available at \url{https://github.com/Lanyu0303/KPO}.
\end{abstract}

\section{Introduction} 

\begin{figure}[t]
    \centering
    \includegraphics[width=1\linewidth]{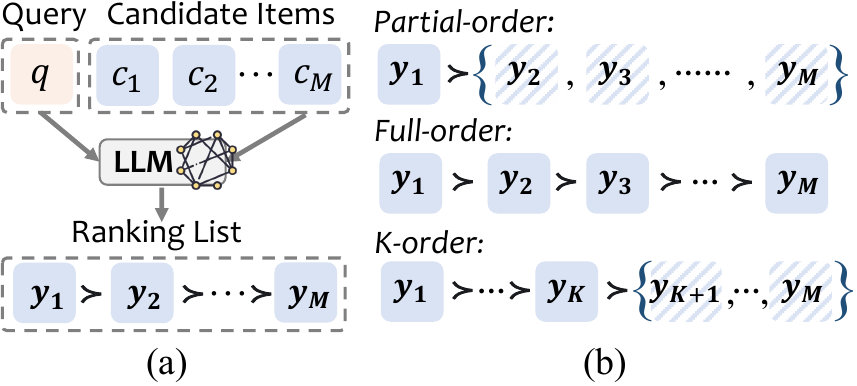}
    \caption{
(a) Illustration of the LLM-based ranking task.
(b) Comparison of three ranking strategies.
    }
    \label{fig:llm_ranking}
\end{figure}

Large Language Models (LLMs) have shown great potential in addressing a wide range of real-world tasks~\cite{llm_survy_application, xu2025personalized}.
By leveraging their semantic reasoning abilities and extensive world knowledge, LLMs can more effectively capture the nuanced relationships between queries and candidate items, making them also promising for ranking tasks~\cite{rankgpt,RankZephyr} — the core of many real-world applications such as product search~\cite{DBLP:journals/corr/abs-2307-03744,DBLP:conf/sigir/Fang0FXNKKA24} and recommendation~\cite{softmaxdpo,llamarec}.  
However, as illustrated in Fig.(\ref{fig:llm_ranking}), ranking tasks extend beyond evaluating the relevance of individual candidates to a user query; they require ranking a list of candidates. 
Yet, LLMs are not explicitly trained to optimize list-wise ranking preferences during pretraining. This limitation has sparked greater research efforts to enhance LLMs' list-wise ranking capabilities~\cite{rankvicuna}.

Among existing approaches, the list-wise Direct Preference Optimization (DPO) method~\cite{dpo, softmaxdpo} has emerged as a promising technique for optimizing LLMs to generate ranked outputs that align with human preferences \textbf{directly at the list level}. 
According to the ranking consistency optimized for a list, existing methods can be categorized into:  
\begin{itemize}[topsep=0pt,itemsep=0pt,parsep=0pt,leftmargin=*]
    \item \textit{Partial-order Method} (e.g., S-DPO~\cite{softmaxdpo}), which simply optimizes the ranking consistency where ``the best item is better than all others,'' i.e., \( y_1{\succ}\) all others. This method focuses on the ranking of the best one, failing to optimize the fine-grained ranking consistency.  
    
    \item \textit{Full-order Method} (e.g., DPO$_{\text{PL}}$~\cite{dpo}), which optimizes complete and fine-grained ranking consistency, i.e., \(y_1{\succ}y_{2}{\succ}\dots\succ\dots\)
    Ideally, this method ensures optimal ranking alignment, but the optimization's inherent difficulties could limit its practical performance.
    
\end{itemize}

Given these, we argue that optimizing top-K ranking consistency would be more appropriate for real-world ranking tasks~\cite{DBLP:journals/informs/AdomaviciusZ16,DBLP:journals/jair/LeL21}. 
In practical scenarios, users typically have limited attention and focus only on the most relevant items, making top-K optimization sufficient to meet their needs. Moreover, this limited attention makes it difficult to obtain accurate preference ranks for less relevant items, rendering ranking optimization for long-tail items inherently unreliable.
Therefore, we propose top-K order ranking preference alignment for LLMs---optimizing the model to align fine-grained ranking consistency for the top-K items while disregarding it for others, 
(i.e., optimizing \( y_1{\succ}\dots{\succ}{y_K}{\succ}\text{all others} \)), as shown in Fig.(\ref{fig:llm_ranking}).

Towards the top-K order ranking preference alignment, we propose \textit{\textbf{K}-order Ranking \textbf{P}reference \textbf{O}ptimization} (KPO). 
The core idea is to extend existing DPO methods' Plackett-Luce preference model~\cite{pl1}, originally designed for full rankings, to accommodate top-K rankings.
Intuitively, KPO works by increasing the relative log probability of \textbf{each} top-K item over all its subsequent items, ensuring both the fine-grained order among the top-K items and the order between the top-K items and the others.
As discussed, KPO is expected to outperform full-order methods due to its closer alignment with real-world scenarios. Additionally, theoretical analysis demonstrates that KPO surpasses existing partial-order methods.

Taking it a step further, in real-world scenarios, the number of most relevant items can vary across queries.
To address this, we extend KPO to handle varying \( K \) values across samples, incorporating a strategy to adaptively determine \( K \) based on LLM confidence in assessing item relevance.
Furthermore, to accommodate varying \( K \), we incorporate a curriculum learning strategy into KPO to simplify the learning process.  
Specifically, we guide KPO to focus on K-order optimization progressively, starting from smaller \( K \) and gradually increasing to larger \( K \). This approach is motivated by the fact that higher \( K \) introduces greater learning challenges, as it requires distinguishing more complete and fine-grained rankings.

The main contributions of this work can be summarized as follows:  
\begin{itemize}[topsep=0pt,itemsep=0pt,parsep=0pt,leftmargin=*]
\item We propose optimizing top-K ranking consistency for LLM ranking preference alignment to better match real-world needs and constraints.

\item We propose KPO for top-K order ranking alignment, incorporating an adaptive strategy to determine suitable \( K \) values for different samples and a curriculum learning strategy to enhance training effectiveness.

\item Extensive experimental results validate KPO's effectiveness while showcasing its high sample efficiency and robustness to noisy logits.

\end{itemize}

\section{Related Work} 
In this section, we delve into related studies from two perspectives: LLM-based ranking and preference alignment in LLMs.

\subsection{LLM-based Ranking}
With the rise of LLMs with strong reasoning abilities, researchers have increasingly explored their potential in ranking tasks~\cite{rankgpt,prp,rankllama,llamarec}. Studies in this area generally follow two approaches: zero-shot usage~\cite{setwise, tourrank} or fine-tuning for enhanced performance~\cite{rankgpt,listt5, rankllama}. In the zero-shot setting, methods like RankGPT~\cite{rankgpt} leverage ChatGPT~\cite{chatgpt,gpt4} to rank candidate passages based on a query. Fine-tuned models, such as RankLLaMA~\cite{rankllama}, use point-wise training to estimate relevance scores, improving reranking precision. LlamaRec~\cite{llamarec} further extends this by introducing a two-stage framework with a verbalizer-based method for generating probability distributions over candidate items. These advancements highlight the growing role of LLMs in ranking tasks, particularly for search and recommendation applications \cite{gao2025sprec,gao2025flower}.

\subsection{Preference Alignment in LLMs}
Preference alignment helps LLMs differentiate between ``good'' and ``bad'' answers using human-labeled data~\cite{rlhf}. 
For example, DPO~\cite{dpo} fine-tunes LLMs with pair-wise preference data, while KTO~\cite{kto}, inspired by Kahneman-Tversky's prospect theory~\cite{kt2}, simplifies this process by utilizing point-wise labels. 
However, both approaches face limitations in effectively handling ranking tasks that require aligning LLMs with multi-item ranking information.
Extensions such as DPO$_{\text{PL}}$~\cite{dpo} and S-DPO~\cite{softmaxdpo} adapt DPO for list-wise settings: DPO$_{\text{PL}}$ targets full-order rankings, while S-DPO handles partial-order rankings. Nonetheless, these methods overlook $K$-order ranking, a critical aspect of ranking tasks.
\section{Problem Definition}
Consider a ranking dataset $\mathcal{D}$ comprising query-candidate pairs, where each $k$-th instance $(q^{(k)}, \mathcal{C}^{(k)}) \in \mathcal{D}$ consists of:  
(1) A query $q^{(k)}$ representing an information need (e.g., search query, recommendation context).
(2) A candidate set $\mathcal{C}^{(k)} = \{c_1^{(k)}, c_2^{(k)}, \dots, c_M^{(k)}\}$ containing $M$ items to be ranked.

The LLM takes as input a concatenated sequence $x^{(k)} = (q^{(k)}, \mathcal{C}^{(k)})$ and aims to generate a permutation $\mathcal{Y}^{(k)} = \{y_1^{(k)} \succ y_2^{(k)} \succ \dots \succ y_M^{(k)}\}$, where $\forall y_i^{(k)} \in \mathcal{C}^{(k)}$, the symbol $\succ$ represents a pair-wise preference relationship.
When ambiguity is absent, we omit the superscript ${}^{(k)}$ for notational simplicity (e.g., $y_i$ instead of $y_i^{(k)}$).

We instantiate this framework through two representative ranking applications:
\begin{itemize}[topsep=0pt,itemsep=0pt,parsep=0pt,leftmargin=*]
    \item \textit{Sequential Recommendation:} The query $q \triangleq [v_1, v_2, \dots, v_m]$ encodes a user's interaction history, where $v_j$ denotes the $j$-th consumed item. 
    \item \textit{Product Search:} The query $q$ represents a textual search intent (e.g., ``wireless noise-canceling headphones'').
\end{itemize}

Both tasks share the core challenge of learning context-aware preference relations, but differ fundamentally in their query semantics - making them ideal testbeds for evaluating the generalization of ranking frameworks.

\section{Methodology}
We first review foundational work in preference modeling to establish the necessary background. Then, we introduce the proposed model in detail.

\subsection{Preliminary}

\noindent\textbf{Preference Modeling.} 
Preference modeling aims to learn a function that captures human preferences over a set of candidate items, enabling applications such as recommender systems, information retrieval, and human-AI alignment. 
One common approach is the Bradley-Terry (BT) model~\cite{bt}, which provides a probabilistic framework for pair-wise preference learning, defining the likelihood of selecting $y_1$ over $y_2$ given context $x$ as:
\begin{equation}
\hat{p}(y_1 \succ y_2 \mid x) = \frac{\exp(r(x, y_1))}{\exp(r(x, y_1)) + \exp(r(x, y_2))},
\label{eq:BT}
\end{equation}
where $r(x, y)$ is a task-specific reward function that quantifies the relative preference for candidate $y$ in context $x$. 
To learn a policy model that aligns with preferences, a widely adopted approach is Direct Preference Optimization (DPO)~\cite{dpo}. DPO formulates the reward function in terms of the policy model $\pi_\theta$ and a reference model $\pi_\text{ref}$:
\begin{equation}
\small
\begin{aligned}
 r(x, y) = \beta \log \frac{\pi_\theta(y \mid x)}{\pi_\text{ref}(y \mid x)} + \beta \log Z(x), 
 \label{eq:reward}
 \end{aligned}
 \end{equation}
where $\beta$ controls the divergence between $\pi_\theta$ and $\pi_{\mathrm{ref}}$. The partition function $Z(x)$ is defined as:
 \begin{equation}
 \small
 Z(x) = \sum_y \pi_{\text{ref}}(y \mid x) \exp\left(\frac{1}{\beta} r(x, y)\right).
 \end{equation}
 
\medskip\noindent\textbf{Full-order Preference Modeling.} 
While the pairwise BT model in Eq.~\eqref{eq:BT} is effective for binary comparisons, it struggles with ranking tasks involving multiple candidate items. To address this limitation, prior work (e.g., DPO$_{\text{PL}}$~\cite{dpo}) has generalized BT to the \textit{list-wise} Plackett-Luce (PL) model~\cite{pl1}, which represents rankings as a full-order sequence $y_1 \succ y_2 \succ \dots \succ y_M$:
\begin{equation}
\small
\begin{aligned}
\hat{p}(y_1 \succ y_2 \succ \dots \succ y_M \mid x) = \prod_{i=1}^{M-1} \frac{\exp(r(x, y_i))}{\sum_{j=i}^M \exp(r(x, y_j))}.
\label{eq:pl_model_pr}
\end{aligned}
\end{equation}

However, full-order methods risk overemphasizing irrelevant item relationships, making optimization more challenging.

\medskip\noindent\textbf{Partial Preference Modeling.} 
To mitigate this, S-DPO~\cite{softmaxdpo} simplifies the PL model by structuring preferences as a single positive candidate against multiple negatives. This modification models preference as $y_1 \succ \{y_2, \dots, y_M\}$:
\begin{equation}
\small
\begin{aligned}
\hat{p}(y_1 \succ \{y_2, \dots, y_M\} \mid x) =  \frac{\exp(r(x, y_1))}{\sum_{j=1}^M \exp(r(x, y_j))}.
\label{eq:s_dpo_model}
\end{aligned}
\end{equation}
While S-DPO reduces computational complexity, it oversimplifies the ranking problem by ignoring nuanced distinctions among top candidates. In real-world applications such as top-K recommendation~\cite{topk_personalized_rec,recranker} and top-K retrieval~\cite{topk_query,topk_retrieval}, users are primarily interested in the relative ordering of the most relevant items. This motivates our proposal for a hybrid approach that combines the strengths of full-order and partial-order models, focusing specifically on accurate top-K preference modeling.

\begin{figure}[htbp]
\centering
\includegraphics[width=1.0\linewidth]{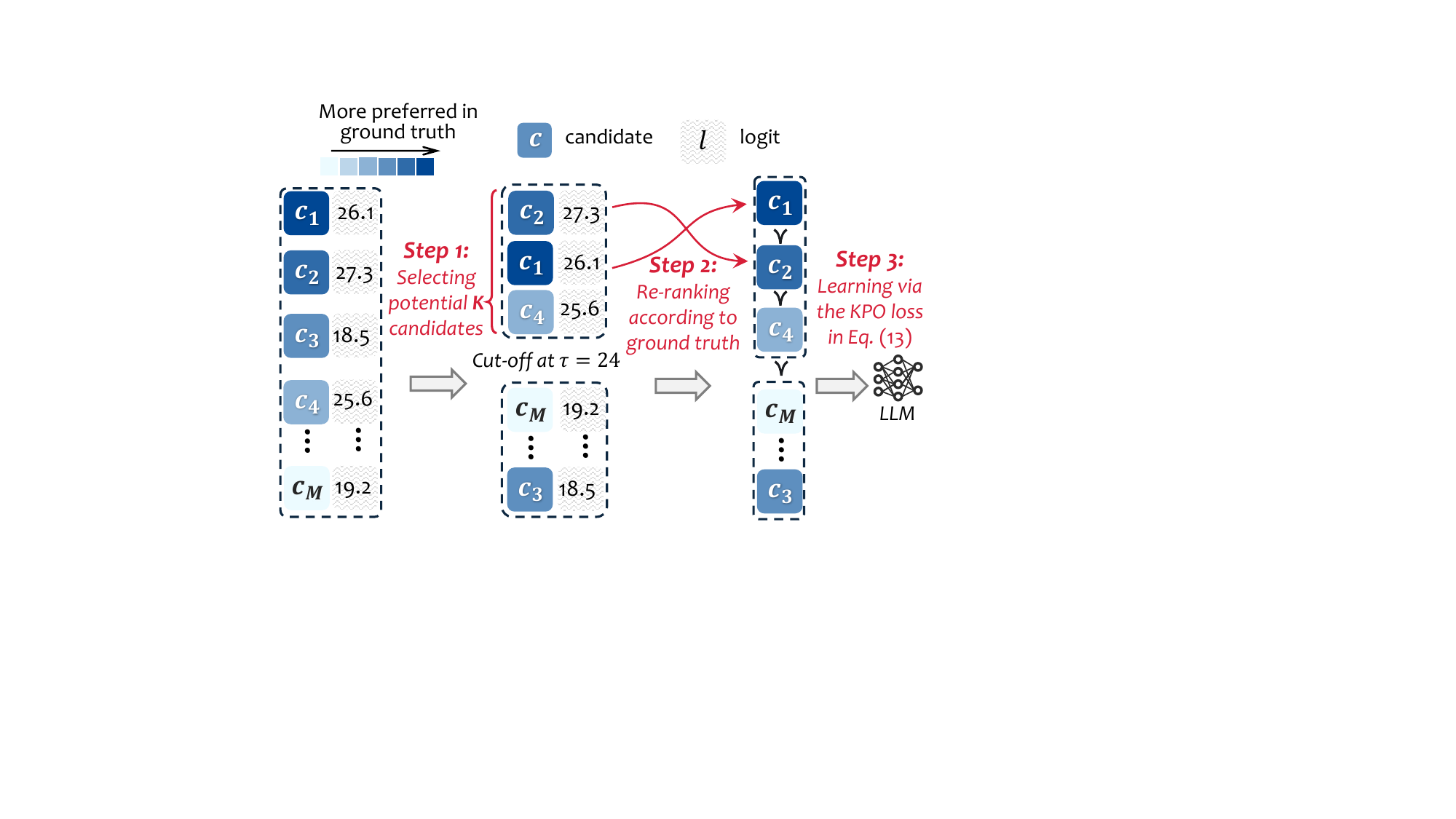}
\caption{
The KPO framework consists of: (1) selecting $K$ candidates via LLM logits above threshold $\tau$, (2) re-ranking top-$K$ candidates to match ground truth, and (3) training the model with the KPO loss in Eq.~\eqref{eq:kpo}. 
}
\label{fig:pipeline}
\end{figure}
\subsection{KPO}
Our goal is to derive a K-order preference:
$y_1 \succ \dots \succ y_K \succ \{y_{K+1}, \dots, y_M$\}
where $y_i \in \mathcal{C}$, $\{y_1, y_2, \dots, y_K\}$ correspond to the top-K relevant items, and $\{y_{K+1}, \dots, y_M\}$ represent the remaining irrelevant items.

Based on the PL model in Eq.~\eqref{eq:pl_model_pr}, we can define the $K$-order preference model as below:
\begin{equation}
\small
\begin{aligned}
&\hat{p}(y_1 \succ \dots \succ y_K \succ \{y_{K+1}, \dots, y_M\} \mid x) \\
&= \prod_{i=1}^{K} \frac{\exp(r(x, y_i))}{\sum_{j=i}^M \exp(r(x, y_j))}.
\label{eq:k_model}
\end{aligned}
\end{equation}
Due to space constraints, the detailed derivation of Eq.~\eqref{eq:k_model} is provided in Appendix~\ref{sec:p_model}.

\medskip\noindent\textbf{Remark:} 
The proposed $K$-order preference framework generalizes existing approaches, with the DPO, DPO$_{\text{PL}}$, and S-DPO emerging as special cases of Eq.~\eqref{eq:k_model}.
When $M = 2$ and $K = 1$, it reduces to DPO's pair-wise preference modeling.
When $K = M$, it recovers DPO$_{\text{PL}}$'s full-order ranking.
When $K = 1$, it simplifies to S-DPO's partial-order formulation.
\medskip

By following the implementation of the reward function $r(x,y)$ from Eq.~\eqref{eq:reward} in DPO, we can derive the loss function $\mathcal{L}_{\rm KPO}$ to maximize $\hat{p}$ on a ranking dataset $\mathcal{D}$ as follows:

\begin{equation}
\small
\begin{aligned}
&\mathcal{L}_{\rm KPO}(\pi_\theta;\pi_{\rm ref}) 
=
-\mathbb{E}_{ (x,y_1,\dots, y_M)\sim\mathcal{D}}\Bigg[
\sum_{i=1}^{K} {\rm log}\,\sigma \Bigg(
-\\ &{\rm log}\,\sum_{j=i+1}^M
{\rm exp}\Bigg(
\beta\,{\rm log}\,\frac{\pi_\theta(y_j|x)}{\pi_{\rm ref}(y_j|x)} 
-\beta\,{\rm log}\,\frac{\pi_\theta(y_i|x)}{\pi_{\rm ref}(y_i|x)}
\Bigg)\Bigg)\Bigg].
\label{eq:kpo}
\end{aligned}
\end{equation}

\medskip\noindent\textbf{Theoretical Analysis:} 

\label{sec:theoretical_ana}
We analyze the optimal top-$K$ ranking accuracy of KPO through the following theorem.

\begin{theorem}
\label{th:th1}
Let $\pi^*$ be the optimal policy that maximizes the KPO objective.
Given a dataset of aggregated preferences $\mathcal{D}_p = \{(x, y_1 \succ \cdots \succ y_K \succ \{y_{K+1}, \ldots, y_M\}\}$.
Assume $\mathcal{D}_p$ contains
ground-truth ranking probabilitie following the PL model. Specifically, for any item $y_i$ and the subset of remaining items $\{y_{i+1}, \dots, y_M\}$, the ranking probability is defined as follows:
\begin{equation}
    \alpha(x, y_i, y_{> i}) = \mathbb{P}(y_i \succ \{ y_{i+1}, \cdots, y_M\}).
\end{equation}
The top-K ranking accuracy of $\pi^*$ is given by:
\begin{equation}
\small
\begin{aligned}
&\mathcal{R}_{\text{KPO}}^*(\mathcal{D}_p, \pi_{\mathrm{ref}}) \\&= \mathbb{E}_{ (x,y_1,\dots, y_M)\sim\mathcal{D}_p} \left[ \prod_{l=1}^{K} \prod_{k=l+1}^{M} \mathbb{I} \left[ \frac{w_l \pi_{\mathrm{ref}}(y_l \mid x)}{w_k \pi_{\mathrm{ref}}(y_k \mid x)} > 1 \right] \right],
\end{aligned}
\end{equation}
where $\frac{w_l}{w_k}$ is defined as 
\begin{equation}
\small
\begin{aligned}
    \frac{w_l}{w_k} = & \Bigg(\frac{\alpha(x, y_l, y_{> l})}{\alpha(x, y_k, y_{> k})} \Bigg)^{1/\beta} \cdot \prod_{i=l}^{k-1}(1-\alpha(x,y_i,y_{> i}))^{-1/\beta}.
\end{aligned}
\label{eq:kpo_wk}
\end{equation}

\end{theorem}
\noindent
The proof is deferred to Appendix~\ref{sec:proof_theorem}.

\medskip
According to Theorem~\ref{th:th1}, we can derive the optimal accuracy of S-DPO as:
\begin{equation}
\small
\begin{aligned}
&\mathcal{R}_{\text{S-DPO}}^*(\mathcal{D}_p, \pi_{\mathrm{ref}}) 
\\&= \mathbb{E}_{ (x,y_1,\dots, y_M)\sim\mathcal{D}_p} \left[ \prod_{l=1}^{K} \prod_{k=l+1}^{M} \mathbb{I} \left[ \frac{w_l' \pi_{\mathrm{ref}}(y_l \mid x)}{w_k' \pi_{\mathrm{ref}}(y_k \mid x)} > 1 \right] \right],
\end{aligned}
\end{equation}
where $\frac{w_l'}{w_k'}$ is defined as 
\begin{equation}
\small
\begin{aligned}
\frac{w_l'}{w_k'} =& \left(\frac{\alpha(x, y_l, y_{> l})}{\alpha(x, y_k, y_{> k})} \right)^{1/\beta} \cdot \prod_{i=l}^{k-1}(1-\alpha(x,y_i,y_{> i}))^{-1/\beta} \\ 
& \cdot \mathbb{I}[l=1] + \mathbb{I}[l\neq 1].
\label{eq:sdpo_wk}
\end{aligned}
\end{equation}

Based on Eq.~\eqref{eq:kpo_wk} and Eq.~\eqref{eq:sdpo_wk}, we can conclude that: \(\frac{w_l}{w_k} > \frac{w_l'}{w_k'}\) for all \( l \in \{2, \dots, K\} \) and \( k \in \{l + 1, \dots, M\} \).
Therefore, we have $\mathcal{R}_{\text{KPO}}(\mathcal{D}_p, \pi_{\text{ref}}) > \mathcal{R}_{\text{S-DPO}}^*(\mathcal{D}_p, \pi_{\text{ref}})$, implying that the optimal ranking accuracy of KPO is greater than S-DPO.
The detailed derivation is provided in Appendix~\ref{sec:app_kpo_sdpo_proof}.

\subsection{Query-adaptive KPO}
In real-world ranking scenarios, the number of relevant candidates $K$ often varies significantly across queries. For instance, a query like ``NVIDIA A40 GPU'' typically has a single authoritative result, while ``budget wireless headphones'' may involve multiple comparable options. To address this, we propose a query-adaptive extension of KPO that dynamically adjusts to each query's characteristics.

\subsubsection{Query-adaptive KPO Loss}
The key challenge lies in determining the appropriate $K$ for each input $x = (q, \mathcal{C})$. We formalize this through a query-adaptive function $\mathcal{K}(x)$ that predicts the number of relevant candidates for a given query. This allows us to extend the KPO loss to its query-adaptive form:
\begin{equation}
\small
\begin{aligned}
&\mathcal{L}_{\rm KPO}^{\mathcal{K}(x)}(\pi_\theta;\pi_{\rm ref})  =
-\mathbb{E}_{ (x,y_1,\dots, y_M)\sim\mathcal{D}}\Bigg[
\sum_{i=1}^{\mathcal{K}(x)} {\rm log}\,\sigma \Bigg(
-\\ &{\rm log}\,\sum_{j=i+1}^M
{\rm exp}\Bigg(
\beta\,{\rm log}\,\frac{\pi_\theta(y_j|x)}{\pi_{\rm ref}(y_j|x)} 
-\beta\,{\rm log}\,\frac{\pi_\theta(y_i|x)}{\pi_{\rm ref}(y_i|x)}
\Bigg)\Bigg)\Bigg]
\label{eq:kpo_adaptive}
\end{aligned}
\end{equation}

\subsubsection{$K$-aware Curriculum Learning}
To effectively train the query-adaptive loss in Eq.~\eqref{eq:kpo_adaptive}, we propose a $K$-aware curriculum strategy~\cite{cl2}. This approach organizes training instances based on their complexity, where complexity is defined by the number of relevant candidates $K$. We treat queries with smaller $K$ values as ``simple samples'', as they require the model to focus on only a few relevant items. Conversely, queries with larger $K$ values are considered ``challenging samples'', demanding more complex ranking decisions.

Following this intuition, we sort the training data in ascending order of $K$, allowing the model to first learn from simpler queries before progressively handling more complex ones. This structured training not only facilitates smoother convergence but also ensures consistent $K$ values within each batch, improving training stability.

\subsubsection{Acquisition of Query-adaptive $K$}
\label{sec:query_adaptive_k}
To determine query-adaptive $K$ values for each input $x = (q, \mathcal{C})$, we leverage the output information of the LLM itself to select $K$ relevant candidates, eliminating the need for additional information.
Specifically, we first use the reference model $\pi_{\mathrm{ref}}$ to compute the logits $logits(q, c_i)$ for each candidate item $c_i \in \mathcal{C}$ based on the given query $q$.
Items with logits exceeding a predefined hyperparameter threshold $\tau$ are regarded as relevant candidates. 
The number of such items is then counted to determine the query-adaptive $K$. Formally, this process is represented as $\mathcal{K}(x)$, defined as:
\begin{equation}
\small
\begin{aligned}
\mathcal{K}(x) = \mathcal{K}(q, \mathcal{C}) = \sum_{i=1}^M \mathbb{I} \left( logits(q, c_i) > \tau \right).
\label{eq:obtain_k}
\end{aligned}
\end{equation}

After obtaining the $K$ values, we generate $K$-order ranking data for KPO training by first sorting candidate items based on their logits to select the top-K items.
These top-K items are then re-ranked using ground truth relevance labels to ensure the correct relative order.
The resulting training data is structured as $y_1 \succ \dots \succ y_{\mathcal{K}(x)} \succ \{y_{\mathcal{K}(x)+1}, \dots, y_M\}$. Details on obtaining the ground truth labels are provided in Appendix~\ref{sec:ground_truth_label}.

The whole pipeline of the proposed method is illustrated in Fig.(\ref{fig:pipeline}).

\subsubsection{Analysis of Time Complexity}
\label{sec:method_time_cost}
The optimization objective of KPO introduces an additional $K$-layer loop compared to S-DPO, which may raise concerns about time complexity. 

To address this potential issue, we conduct an analysis of the time required for the actual optimization process. Specifically, the parameter update process can be divided into three phases:
\begin{itemize}[topsep=0pt,itemsep=0pt,parsep=0pt,leftmargin=*]
\item \textbf{Phase 1}: Compute $M$ ``rewards'' ($r_i = \beta \log \frac{\pi_\theta(y_i|x)}{\pi_{\text{ref}}(y_i|x)}$).
\item \textbf{Phase 2}: Use the rewards to compute the loss.
\item \textbf{Phase 3}: Update model parameters via loss backpropagation.
\end{itemize}

The $K$-layer loop introduced by KPO occurs in Phase 2. However, the actual runtime of Phase 2 is significantly shorter compared to Phase 1 and Phase 3, and thus does not impact the overall runtime of the method.
Detailed experimental results supporting this conclusion are provided in Appendix~\ref{sec:app_time_cost}.

\section{Experiments}
\begin{table*}[htbp]
\centering
\small
 \renewcommand\tabcolsep{3pt} 
 \renewcommand\arraystretch{1.1} 
\begin{tabular}{l|ccccc|ccccc|cc}
\toprule
\multirow{2}{*}{Method} & \multicolumn{5}{c|}{MovieLens} &\multicolumn{5}{c|}{Goodreads}&\multicolumn{2}{c}{Shopping Queries}\\
 &HR@1 & HR@5 & HR@10 & N@5 & N@10 &HR@1 & HR@5 & HR@10 & N@5 & N@10 &N@5 & N@10 \\
\midrule
KTO & 0.5368& 0.8421& 0.9474& 0.6996& 0.7342& 0.4875& 0.8486& 0.9534& 0.6808& 0.7147& 0.7327& 0.7525\\
\midrule
DPO & 0.5263& 0.8632& 0.9579& 0.7052& 0.7348& 0.4908& 0.8569& 0.9584& 0.6858& 0.7216& 0.7356& 0.7531\\
SimPO & 0.5263& \textbf{0.8842}& 0.9579& 0.7217& 0.7448& 0.4842& 0.8569& 0.9551& 0.6794& 0.7113& 0.7392& 0.7560\\
cDPO & 0.5158& 0.8632& \textbf{0.9684}& 0.6960& 0.7290& 0.4509& 0.8536& 0.9534& 0.6651& 0.6979& 0.7321& 0.7503\\
\midrule
S-DPO & 0.5368& 0.8526& 0.9474& 0.7062& 0.7369& 0.4842& 0.8353& 0.9484& 0.6712& 0.7083& 0.7288& 0.7480\\
DPO$_{\text{PL}}$ & 0.5474& 0.8737& 0.9474& 0.7229& 0.7463& 0.4859& 0.8619& \textbf{0.9634}& 0.6876& 0.7205& 0.7363& 0.7529\\
KPO$_{\text{CUT}}$ & 0.5474& 0.8632& \textbf{0.9684}& 0.7167& 0.7493& 0.4992& 0.8453& 0.9468& 0.6852& 0.7182& 0.7347& 0.7521\\
\midrule
KPO & \textbf{0.5579}& \textbf{0.8842}& \textbf{0.9684}& \textbf{0.7361}& \textbf{0.7620}& \textbf{0.5042}& \textbf{0.8719}& 0.9584& \textbf{0.6994}& \textbf{0.7272}& \textbf{0.7477}& \textbf{0.7631}\\
\bottomrule
\end{tabular}
\caption{\textbf{Comparison with preference alignment methods.} Bold indicates the best performance.}
\label{tab:dpo_compare}
\end{table*}
In this section, we aim to answer the following research questions (RQ):
\begin{itemize}[topsep=0pt,itemsep=0pt,parsep=0pt,leftmargin=*]
\item \textbf{RQ1}: How does KPO perform in the recommendation and product search tasks?
\item \textbf{RQ2}:  What are the effects of the key components and hyperparameters?
\item \textbf{RQ3}: How does KPO perform in terms of sample efficiency and robustness to noisy logits?
\end{itemize}
\subsection{Experimental Setup}

We organize experiments on two typical ranking tasks: recommendation and product search.

\subsubsection{Datasets}

\textbf{For the recommendation task}, we utilize the MovieLens~\cite{movielens} and Goodreads~\cite{goodreads1} datasets.
The user interaction sequences in each dataset are chronologically sorted and then split into training, validation, and test sets in an 8:1:1 ratio. 

\textbf{For the product search task}, we used the Shopping Queries dataset~\cite{esci}, which includes queries paired with up to 40 candidate products.
Each product is assigned a four-level score ($\{0, 1, 2, 3\}$) representing its relevance to the query, which can serve as the ground truth label.
Queries are grouped and randomly split into training, validation, and test sets in an 8:1:1 ratio.

The detailed description of the datasets and their statistical information is provided in Appendix~\ref{sec:app_datasets}.

\subsubsection{Evaluation Setting}
We evaluate the model's ability to rank 20 candidate items based on a given query.

\textbf{For the recommendation task}, the ground truth item is the user's most recently interacted item. The candidate list includes this ground truth item and 19 randomly sampled items. The model's performance is evaluated based on its ability to rank the ground truth item higher, using Hit Ratio (HR@1, 5, 10) and Normalized Discounted Cumulative Gain (N@5, 10).

\textbf{For the product search task}, multiple ground truth items have relevance labels, we evaluate the model using N@5 and N@10 to measure its ability to prioritize highly relevant items. Additional results for the setting with a single ground truth item are provided in Appendix~\ref{sec:esci_single}.

\subsubsection{Implementation Details}

\label{sec:implementation}
Our experiments are conducted on eight NVIDIA A40 GPUs.
We use the Llama-3.2-3B-Instruct \cite{llama32} model as the backbone. 
In the supervised fine-tuning (SFT) stage, the model is trained for 5 epochs with a learning rate of 1e-4. In the preference alignment stage, the learning rate is reduced to 1e-5, and training is performed over 3 epochs. The global batch size is fixed at 128.
Refer to Appendix~\ref{sec:app_implemetation} for more implementation details.

\subsection{Overall Performance (RQ1)}
In this section, we compare KPO with other preference alignment methods and non-preference alignment methods to evaluate the effectiveness of KPO.
\begin{table}[htbp]
\centering
\renewcommand\arraystretch{1.3} 

 \setlength{\tabcolsep}{0.5mm}{
 \resizebox{\linewidth}{!}{
\begin{tabular}{lll}
\toprule
Method & Modeling & Objective \\
\toprule
KTO & $y$ & $\lambda_y - v_{\text{KTO}}(x, y)$ \\
\midrule
DPO & $y_1 \! \succ \! y_2$ & $-\! \log \sigma \left(r_1 - r_2\right)$ \\
SimPO & $y_1 \! \succ \! y_2$ & $-\log \!  \sigma \! \left(\! \frac{\beta}{|y_1|} \! \log\! \pi_\theta(y_1|x) \!-\! \frac{\beta}{|y_2|} \log \pi_\theta(y_2|x) \!-\! \gamma \! \right)$ \\
cDPO & $ y_1 \gtrless y_2$ & $-(1-\epsilon)\log \sigma\left( r_1-r_2\right) - \epsilon \log \sigma\left(r_2 - r_1 \right)$ \\
 \midrule
 S-DPO &$y_1 \! \succ \! \{y_2, \dots, y_M\}$ & ${\rm log}\,\sigma\left(-{\rm log}\,\sum_{j=2}^M{\rm exp}\left(r_j - r_1\right)\right)$\\
DPO$_{\text{PL}}$ & $y_1 \! \succ \! y_2 \! \succ \! \dots \! \succ \! y_M$ & $\sum_{i = 1}^{M -1} {\rm log}\,\sigma\left(-{\rm log}\,\sum_{j=i+1}^M{\rm exp}\left(r_j - r_i\right)\right)$\\
 KPO$_{\text{CUT}}$ &$y_1 \! \succ \! y_2 \! \succ \! \dots \! \succ \! y_{\mathcal{K}(x)}$ & $\sum_{i=1}^{\mathcal{K}(x)-1}{\rm log}\,\sigma\left(-{\rm log}\,\sum_{j=i+1}^{\mathcal{K}(x)}{\rm exp}\left(r_j-r_i\right)\right)$ \\
 \midrule
 \multirow{2}{*}{KPO} &$y_1 \! \succ \! y_2 \! \succ \! \dots \! \succ \! y_{\mathcal{K}(x)}$ &  \multirow{2}{*}{$\sum_{i=1}^{\mathcal{K}(x)}{\rm log}\,\sigma\left(-{\rm log}\,\sum_{j=i+1}^M{\rm exp}\left(r_j - r_i\right)\right)$}\\
 & $ \! \succ \! \{y_{\mathcal{K}(x)+1}, \dots y_M\}$ & \\
\bottomrule
\end{tabular}}}
\caption{\textbf{
Modeling approaches and optimization objectives for preference alignment methods.} For convenience, we define $r_i = \beta \log \frac{\pi_\theta(y_i|x)}{\pi_{\text{ref}}(y_i|x)}$.
The detailed definitions of KTO are provided in the Appendix~\ref{sec:kto}.
}
\label{tab:dpo_loss}
\end{table}
\subsubsection{Comparison with Preference Alignment Methods}
\begin{table*}[htbp]
\centering
\small
 \renewcommand\arraystretch{1.2} 

\begin{tabular}{l|ccccc|ccccc}
\toprule
\multirow{2}{*}{Method} & \multicolumn{5}{c|}{MovieLens} &\multicolumn{5}{c}{Goodreads}\\
 &HR@1 & HR@5 & HR@10 & N@5 & N@10 &HR@1 & HR@5 & HR@10 & N@5 & N@10\\
\midrule
SASRec &0.4043 &0.8298 &0.9043 &0.6356 &0.6588 &0.3661 &0.7654 &0.9118 &0.5763 &0.6238\\
GRU4Rec & 0.4526& 0.8316& 0.9053& 0.6498& 0.6738& 0.3478& 0.7504& 0.9251& 0.5606&0.6185\\
Caser & 0.3404& 0.7979& 0.9255& 0.5845& 0.6259& 0.4133& 0.8083& 0.9283& 0.6251&0.6640\\
\midrule
MoRec & 0.2737& 0.6842& 0.8211& 0.4783& 0.5244& 0.3111& 0.7121& 0.8918& 0.5240&0.5824\\
LLaRA & 0.4565& 0.8370& 0.9130& 0.6376& 0.6630& 0.4742& 0.8053& 0.9235& 0.6341&0.6713\\
\midrule
RankGPT$_\text{3.5}$ & 0.2211& 	0.5579& 	0.7368& 	0.3920& 	0.4506& 	0.3389& 	0.5763& 	0.7288& 	0.4674& 	0.5158 \\
LlamaRec & 0.5158&	0.8526&	0.9474&	0.6999&	0.7402&	0.4842&	0.8419&	0.9501&	0.6765&	0.7114 \\
\midrule
SFT & 0.5053& 0.8526& 0.9368& 0.6983& 0.7255& 0.4809& 0.8369& 0.9468& 0.6675&0.7034\\
KPO$_{\text{CL}}$ & \textbf{0.5684}& \textbf{0.8947}& \textbf{0.9684}& \textbf{0.7381}& \textbf{0.7637}& \textbf{0.5158}& \textbf{0.8735}& \textbf{0.9667}& \textbf{0.7024}&\textbf{0.7353}\\
\bottomrule
\end{tabular}

\caption{\textbf{Comparison with other recommendation models and rankers.} Bold indicates the best performance.}
\label{tab:rec_compare}
\end{table*}
To evaluate the effectiveness of KPO loss, we compare it with various preference alignment methods on the recommendation and product search tasks.

\textbf{Baselines.}
We compare KPO to various baselines, including KTO~\cite{kto}, DPO~\cite{dpo}, SimPO~\cite{simpo}, Conservative DPO (cDPO)~\cite{cdpo}, S-DPO~\cite{softmaxdpo}, and DPO$_{\text{PL}}$~\cite{dpo}. We also introduce KPO$_\text{CUT}$, a KPO variant that cuts off tail-irrelevant items $\{y_{\mathcal{K}(x)+1}, \dots, y_{M}\}$ for comparison. Objective formulations are summarized in Table~\ref{tab:dpo_loss}, with detailed baseline descriptions in Appendix~\ref{sec:app_preference_baseline}.

\textbf{Results.}
The experimental results are summarized in Table~\ref{tab:dpo_compare}. To fairly evaluate the loss function's effectiveness, KPO's performance is reported without curriculum learning.  
The key findings are as follows:  
(1) KPO consistently outperforms other methods across most metrics, demonstrating its effectiveness.  
(2) KPO surpasses KPO$_{\text{CUT}}$, highlighting the importance of irrelevant items in helping the model distinguish between relevant and irrelevant ones.  
(3) Although KPO slightly underperforms DPO$_{\text{PL}}$ in HR@10 on the Goodreads dataset, the HR@10 values across all methods are already high. Notably, KPO achieves a higher N@10 than DPO$_{\text{PL}}$, reflecting better overall ranking quality. 

\subsubsection{Comparison with Non-Preference Alignment Methods}
To verify whether KPO outperforms other non-preference alignment methods, this section focuses on the recommendation task and compares KPO with various recommendation models and rankers.

\textbf{Baselines.}
We thoroughly compare KPO with three categories of models: traditional recommendation models (SASRec~\cite{sasrec}, GRU4Rec~\cite{gru4rec}, Caser~\cite{caser}), LLM-based recommendation models (MoRec~\cite{morec}, LLaRA~\cite{llara}) and LLM-based rankers (RankGPT$_{\text{3.5}}$~\cite{rankgpt}, LlamaRec~\cite{llamarec}).
The detail description of the models can be found in Appendix~\ref{sec:app_rec_models} and~\ref{sec:app_rec_models}.

\textbf{Results.}
We evaluate the full KPO method with $K$-aware curriculum learning (KPO$_{\text{CL}}$) against baseline models, including the SFT model for comparison.
As shown in Table~\ref{tab:rec_compare}, KPO$_{\text{CL}}$ significantly outperforms baseline models, demonstrating its effectiveness.
This improvement likely stems from the fact that baseline models are trained based on single ground truth items, neglecting the ranking relationships among multiple items, a core focus of the KPO method.

\begin{figure}[htbp]
\centering
\includegraphics[width=1.0\linewidth]{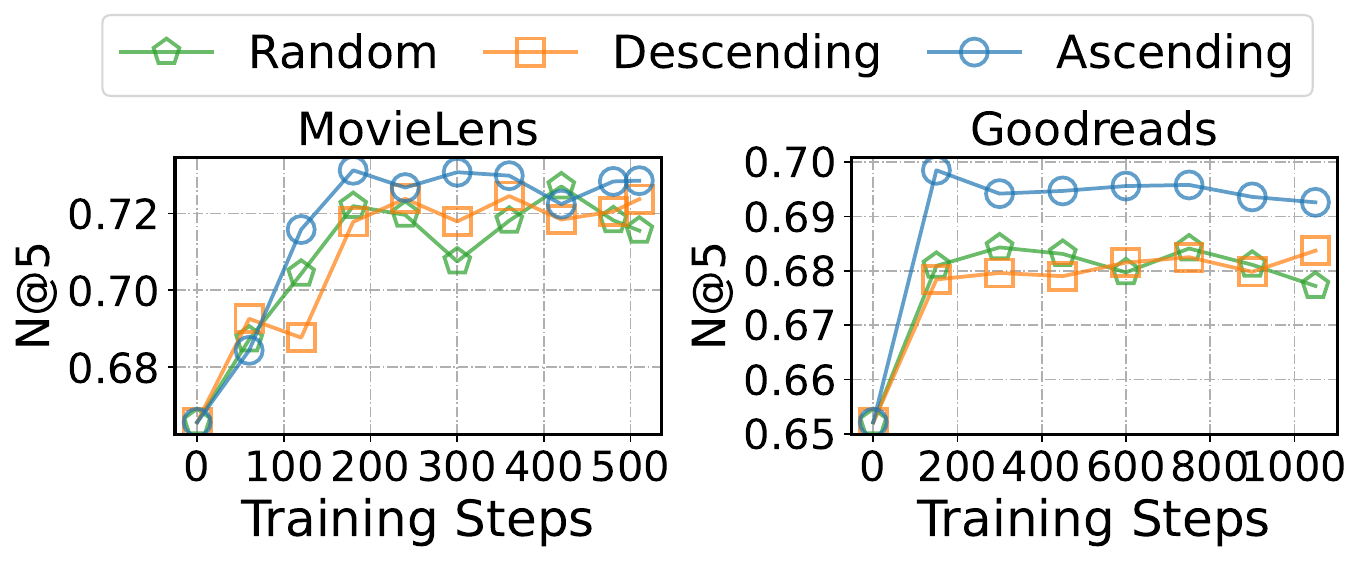}
\caption{N@5 on MovieLens and Goodreads validation set under different training orders.}
\label{fig:cl}
\end{figure}
\subsection{Effectiveness of Key Components (RQ2)}
We investigate the effects of the following key components of our method: (1) $K$-aware curriculum learning, (2) query-adaptive $K$, (3) $\beta$ in Eq.~\eqref{eq:kpo_adaptive}, and (4) threshold $\tau$ in Eq.~\eqref{eq:obtain_k}.
\subsubsection{$K$-aware Curriculum Learning}
To verify the effectiveness of $K$-aware curriculum learning, we design three training orders for the datasets:  
(1) ``Random'': The dataset is randomly shuffled.
(2) ``Descending'': The dataset is sorted by $K$ in descending order.  
(3) ``Ascending'': The dataset is sorted by $K$ in ascending order.  

We evaluate the impact of training orders by plotting N@5 curves on the validation set (Fig.(\ref{fig:cl})). The results show that the ``Ascending'' order outperforms ``Random'' and ``Descending'' orders in both overall performance and stability, underscoring the effectiveness of $K$-aware curriculum learning. Test set results are provided in Appendix~\ref{sec:app_cl}.

\begin{table}[htbp]
\centering
\small
 \renewcommand\tabcolsep{3pt} 
 \renewcommand\arraystretch{1.2} 
\begin{tabular}{c|ccccc}
\toprule
\multirow{2}{*}{$K$} & \multicolumn{5}{c}{MovieLens} \\
 &HR@1 & HR@5 & HR@10 & N@5 & N@10  \\
\midrule
1 & 0.5368& 0.8526& 0.9474& 0.7062& 0.7369\\
3 & 0.5579& 0.8737& 0.9684& 0.7279& 0.7574\\
5 & 0.5474& 0.8737& 0.9684& 0.7251& 0.7526\\
7 & 0.5474& 0.8632& 0.9579& 0.7258& 0.7469\\
10 & 0.5474& 0.8737& 0.9474& 0.7229& 0.7463\\
\midrule
query-adaptive & \textbf{0.5579}& \textbf{0.8842}& \textbf{0.9684}& \textbf{0.7361}& \textbf{0.7620}\\
\bottomrule
\end{tabular}
\caption{\textbf{Comparison of query-adaptive and fixed $K$.}}
\label{tab:fix_k}
\end{table}
\subsubsection{Query-adaptive $K$}
To evaluate the effectiveness of query-adaptive $K$, we compare the performance of query-adaptive KPO against KPO with fixed $K$ values ($[1, 3, 5, 7, 10]$).
As shown in Table~\ref{tab:fix_k}, query-adaptive $K$ consistently outperforms fixed $K$, highlighting its effectiveness.

\subsubsection{$\beta$ in the Loss Function Eq.~\eqref{eq:kpo_adaptive}}  
Fixing $\tau$ at 24, the $\beta$ is varied across $[0.1, 0.5, 1.0, \\3.0, 5.0]$. Typically, smaller $\beta$ values indicate stronger influence of preference signals on the LLM, while larger values suggest weaker influence. As shown in Fig. (\ref{fig:hyper_merge}), the best performance occurs at $\beta = 1.0$, with higher $\beta$ values leading to a notable drop in HR@1. This underscores the importance of effectively leveraging preference signals in ranking tasks.

\subsubsection{Threshold $\tau$ in Eq.~\eqref{eq:obtain_k}}  
\begin{figure}[htbp]
\centering
\includegraphics[width=1.0\linewidth]{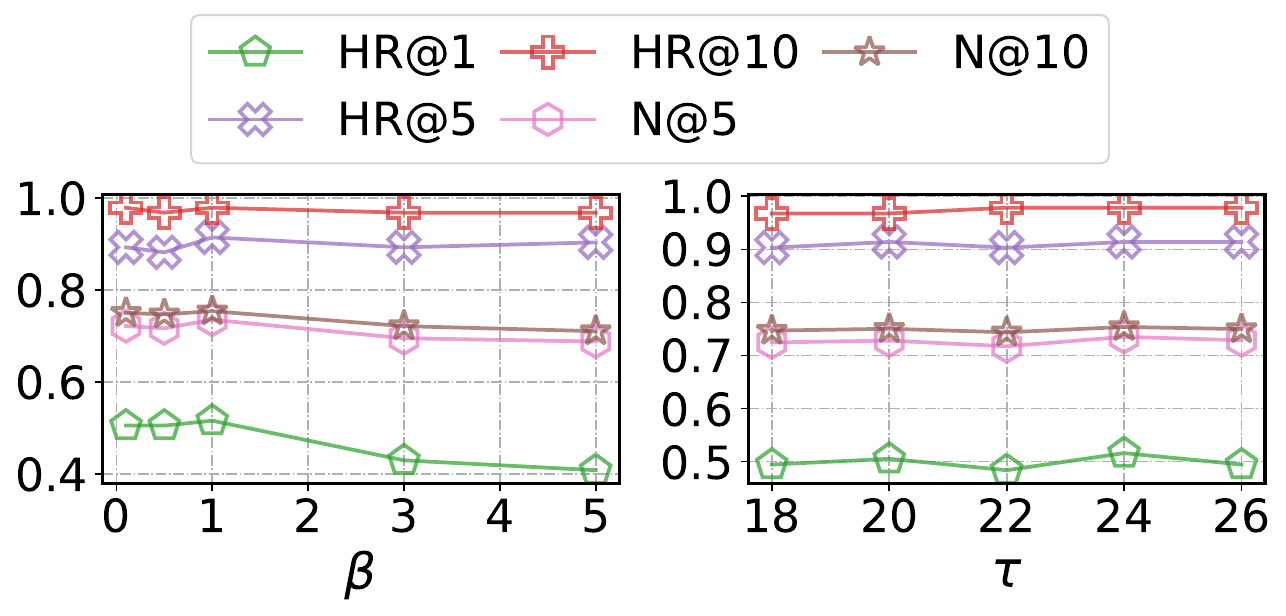}
\caption{Study of $\beta,\tau$ on MovieLens validation set.}
\label{fig:hyper_merge}
\end{figure}

Fixing $\beta$ at 1.0, the $\tau$ is varied across $[18, 20, 22, \\24, 26]$. 
Based on the experimental results in Fig.(\ref{fig:hyper_merge}), we can find that $\tau = 24$ is the optimal value.
Additionally, the performance of the model is not significantly affected by variations in $\tau$, further indicating that KPO demonstrates a certain level of robustness and stability.

\subsection{In-depth Analysis of KPO (RQ3)}
In this section, we conduct an in-depth analysis of KPO from three key perspectives: (1) sample efficiency, and (2) robustness to noisy logits (3) applicability across various backbone models.

\subsubsection{Sample Efficiency}
As shown in Section \S\ref{sec:method_time_cost}, KPO and S-DPO exhibit similar runtimes.
This section highlights how the $K$-layer loop in KPO improves sample efficiency. Fig.(\ref{fig:reward}) compares the reward curves of the top-1 item during training for both methods.

Fig.(\ref{fig:reward}) shows that KPO consistently outperforms S-DPO in reward with the same number of training steps. Moreover, KPO achieves the same reward level as S-DPO in fewer steps, underscoring its superior sample efficiency.

\subsubsection{Robustness to Noisy Logits}
\begin{figure}[htbp]
    \centering
    \subfloat[]{\includegraphics[width=0.5\linewidth]{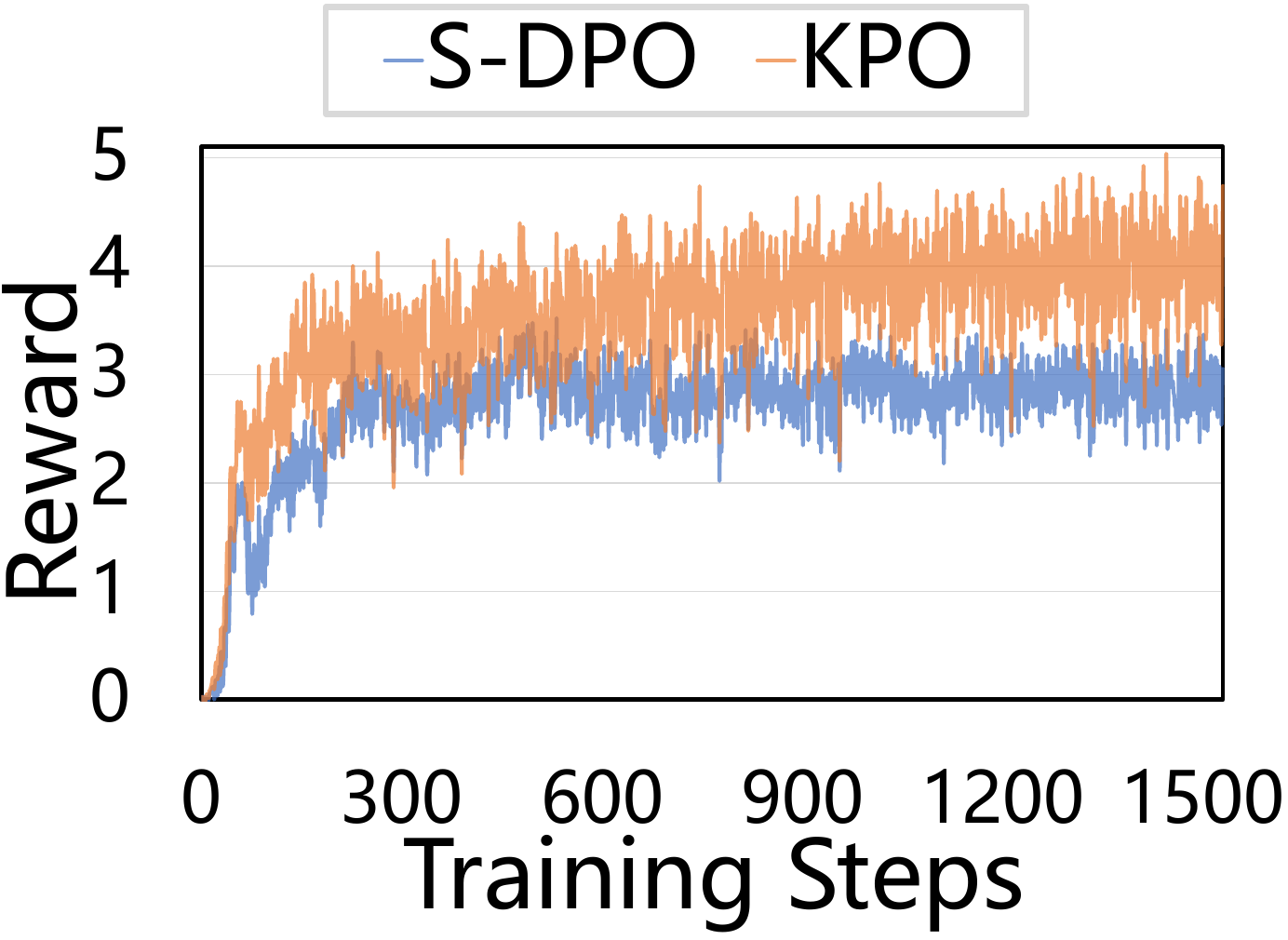}\label{fig:reward}}
    \subfloat[]{\includegraphics[width=0.5\linewidth]{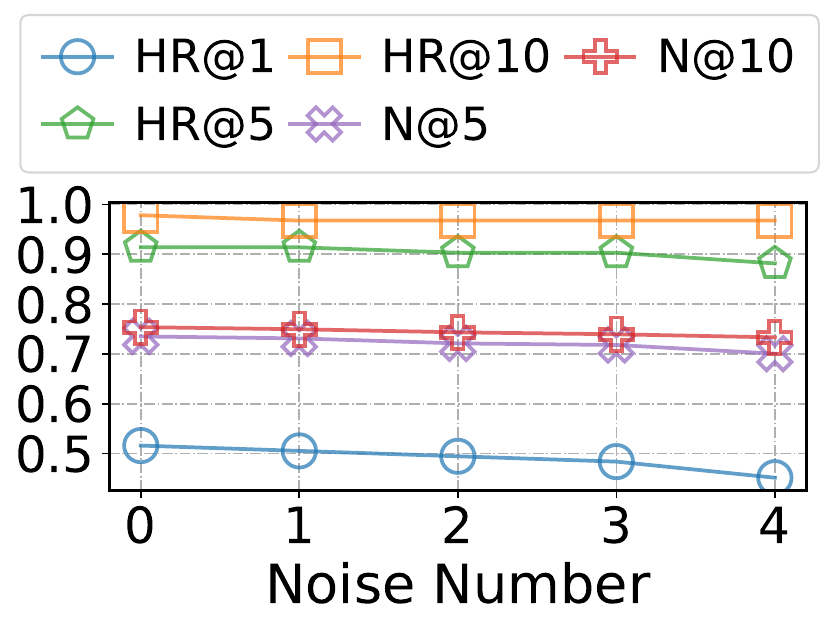}\label{fig:noise}}
    \caption{(a) Comparison of the reward of the top-1 item between KPO and S-DPO on MovieLens.
    (b) Study of noise on the MovieLens validation set.
    }
    \label{fig:reward_noise}
\end{figure}

Since using LLM logits to determine $K$ candidates may introduce inaccuracies, we investigate how such errors impact KPO's training performance.
To simulate these inaccuracies, we introduce a noise-adding mechanism that randomly swaps the logits of two items within $M$ candidates, resulting in false top-K selections.
We then assess performance as the number of swaps increases. Fig.(\ref{fig:noise}) presents the experimental results on the MovieLens validation dataset.
The experiments demonstrate that KPO maintains relatively stable performance despite increasing noise levels, highlighting its robustness to imperfect logit estimates from the LLM.

\subsubsection{Applicability across Various Backbones}
\begin{table*}[htbp]
\centering
 \renewcommand\tabcolsep{5pt} 
 \renewcommand\arraystretch{0.6} 
\begin{tabular}{ll|l|ccccc}
\toprule
\multirow{2}{*}{Model}&\multirow{2}{*}{Size}&\multirow{2}{*}{Method} & \multicolumn{5}{c}{MovieLens} \\
 & & &HR@1 & HR@5 & HR@10 & N@5 & N@10  \\
\midrule
\multirow{4}{*}{Llama-3.2-Instruct}&\multirow{4}{*}{3B}&SFT & 0.5053	&0.8526	&0.9368	&0.6983	&0.7255\\
 & &DPO & 0.5263	&0.8632	&0.9579	&0.7052	&0.7348\\
 & &DPO$_{\text{PL}}$& 0.5474	&0.8737	&0.9474	&0.7229	&0.7463\\
 & &KPO & \textbf{0.5579} & \textbf{0.8842} & \textbf{0.9684} & \textbf{0.7361} & \textbf{0.7620} \\
\midrule
 \multirow{4}{*}{Llama-3.2-Instruct}&\multirow{4}{*}{1B}&SFT & 0.4526&	0.8316&	0.9368	&0.6569&	0.6908\\
 & &DPO & 0.5053	&0.8526&	0.9368&	0.7003&	0.7247\\
 & &DPO$_{\text{PL}}$& 0.5263&	0.8632&	0.9474&	0.7122	&0.7333\\
 & &KPO & \textbf{0.5368} & \textbf{0.8737} & \textbf{0.9579} & \textbf{0.7233} & \textbf{0.7401}\\
\midrule
\multirow{4}{*}{Qwen2.5-Instruct}& \multirow{4}{*}{500M}&SFT & 0.4316	&0.8421	&0.9158	&0.6467	&0.6717\\
 & &DPO & 0.4632	&0.8421	&0.9474	&0.6735	&0.7061\\
 & &DPO$_{\text{PL}}$& 0.4842	&0.8526	&0.9474	&0.6782	&0.7086\\
 & &KPO & \textbf{0.5053} & \textbf{0.8632} & \textbf{0.9579} & \textbf{0.6837} & \textbf{0.7198} \\
\midrule
\multirow{4}{*}{SmolLM2-Instruct}& \multirow{4}{*}{135M}&SFT & 0.0526	&0.2105	&0.4842	&0.1228	&0.2137\\
 & &DPO & 0.0632	&0.2526&	0.5053&	0.1935&	0.2603\\
 & &DPO$_{\text{PL}}$& 0.0737	&0.2737	&0.5263	&0.2040	&0.2597\\
 & &KPO & \textbf{0.0947} & \textbf{0.3263} & \textbf{0.5368} & \textbf{0.2120} & \textbf{0.2645}\\
\bottomrule
\end{tabular}
\caption{\textbf{Performance comparison across various models.} Bold indicates the best performance.}
\label{tab:various_models}
\end{table*}
In this section, we investigate whether KPO can consistently improve performance across various backbone models. Due to computational resource constraints, we selected Llama-3.2-3B-Instruct~\cite{llama32}, Llama-3.2-1B-Instruct, Qwen2.5-500M-Instruct~\cite{qwen2.5}, and SmolLM2-135M-Instruct~\cite{allal2025smollm2smolgoesbig} as our experimental models.

The experimental results are presented in Table~\ref{tab:various_models}.
For better comparison, we also report the performance of SFT, DPO, and DPO$_{\text{PL}}$. According to the results, KPO consistently outperforms the other methods across various backbone models, further demonstrating the effectiveness of KPO.

\section{Conclusion} 
In this study, we propose a novel method called KPO, designed to address the limitations of existing approaches that rely on full-order or partial-order ranking but often neglect the significance of top-K ranking.
In detail, we introduce the $K$-order ranking, which prioritizes fine-grained ranking consistency for the top-K items while disregarding less relevant ones.
Building on this foundation, we extend the PL model to accommodate top-K ranking and develop the corresponding KPO loss.
Additionally, we derive a theoretical formula for the optimal accuracy achievable by KPO, thereby theoretically demonstrating that KPO outperforms S-DPO.
Considering the varying number of relevant items across queries, we make KPO query-adaptive, enabling it to dynamically adjust $K$ for each query.
To further improve training efficiency and stability, we introduce $K$-aware curriculum learning, which allows LLMs to progressively learn from simpler to more complex data.
Extensive experiments show that KPO significantly outperforms existing preference alignment methods, highlighting not only the effectiveness of top-K ranking but also the critical role of query-adaptive $K$.
\section*{Limitations}
In this paper, we propose a query-adaptive KPO framework that dynamically determines the $K$-order for candidate items based on each query. While our approach has demonstrated effectiveness in experiments, the method for obtaining the query-adaptive $K$ remains heuristic and does not guarantee that the resulting $K$ is optimal.

On one hand, we rely on the logits generated by the LLM to represent the relevance between candidate items and the query. However, these logits may not always provide an accurate measure of relevance. It will be our future work to investigate more precise methods for assessing relevance.

On the other hand, our approach determines $K$ by counting the number of items whose logits exceed a predefined threshold $\tau$. This highlights that $K$ is highly sensitive to the choice of this hyperparameter. In future work, we will explore strategies to derive a more accurate and optimal $K$.

\section*{Acknowledgement}
This work is supported by the National Natural Science Foundation of China (62272437, 62402470), Anhui Provincial Natural Science Foundation (2408085QF189), the Fundamental Research Funds for the Central Universities of China
(WK2100000053), and the advanced computing resources provided by the Supercomputing Center of the USTC.

\bibliography{custom}

\appendix
\setcounter{theorem}{0} 
\renewcommand{\thetheorem}{1} 
\section{Mathematical Derivation}
\subsection{Preference Modeling Derivation}
\label{sec:p_model}
In this section, we prove that the preference $y_1 \succ \dots \succ y_K \succ \{y_{K+1}, \dots, y_M\}$ can be expressed as:
\begin{equation}
\small
\begin{aligned}
&\hat{p}(y_1 \succ \dots \succ y_K \succ \{y_{K+1}, \dots, y_M\} \mid x) \\
&= \prod_{i=1}^{K} \frac{\exp(r(x, y_i))}{\sum_{j=i}^M \exp(r(x, y_j))}.
\label{eq:k_model_app}
\end{aligned}
\end{equation}

Specifically, based on Eq.~\eqref{eq:pl_model_pr}, we can derive step by step as follows:
\begin{equation}
\small
\begin{aligned}
&\hat{p}(y_1 \succ \dots \succ y_K \succ y_{K+1}, y_{K+2}, \dots, y_M \mid x) \\
&= \sum_{\text{Per}(y_{K+1}, \dots, y_M)} \prod_{i=1}^{M-1} \frac{\exp(r(x, y_i))}{\sum_{j=i}^M \exp(r(x, y_j))} \\
&= \prod_{i=1}^{K} \frac{\exp(r(x, y_i))}{\sum_{j=i}^M \exp(r(x, y_j))} \times \\&  \sum_{\text{Per}(y_{K+1}, \dots, y_M)} \prod_{i=K+1}^{M-1} \frac{\exp(r(x, y_i))}{\sum_{j=i}^M \exp(r(x, y_j))} \\
&= \prod_{i=1}^{K} \frac{\exp(r(x, y_i))}{\sum_{j=i}^M \exp(r(x, y_j))} \times \\& \sum_{\text{Per}(y_{K+1}, \dots, y_M)} p(y_{K+1} \succ \dots \succ y_M \mid x) \\
&= \prod_{i=1}^{K} \frac{\exp(r(x, y_i))}{\sum_{j=i}^M \exp(r(x, y_j))}, \\
\label{eq:k_model_full}
\end{aligned}
\end{equation}
where $\text{Per}(y_{K+1}, \dots, y_M)$ denotes the set of all permutations of $y_{K+1}, \dots, y_M$.

\subsection{Proof of the Ranking Accuracy Theorem}
\label{sec:proof_theorem}
In this section, we provide the proof of the theorem presented in Section~\S\ref{sec:theoretical_ana}, building on the method outlined in~\cite{ranking_proof}.
\begin{theorem}
Let $\pi^*$ be the optimal policy that maximizes the KPO objective.
Given a dataset of aggregated preferences $\mathcal{D}_p = \{(x, y_1 \succ \cdots \succ y_K \succ \{y_{K+1}, \ldots, y_M\}\}$.
Assume $\mathcal{D}_p$ contains
ground-truth ranking probabilitie following the PL model. Specifically, for any item $y_i$ and the subset of remaining items $\{y_{i+1}, \dots, y_M\}$, the ranking probability is defined as follows:
\begin{equation}
    \alpha(x, y_i, y_{> i}) = \mathbb{P}(y_i \succ \{ y_{i+1}, \cdots, y_M\})
\end{equation}
The top-K ranking accuracy of $\pi^*$ is given by:
\begin{equation}
\small
\begin{aligned}
&\mathcal{R}_{\text{KPO}}^*(\mathcal{D}_p, \pi_{\mathrm{ref}}) \\&= \mathbb{E}_{ (x,y_1,\dots, y_M)\sim\mathcal{D}_p} \left[ \prod_{l=1}^{K} \prod_{k=l+1}^{M} \mathbb{I} \left[ \frac{w_l \pi_{\mathrm{ref}}(y_l \mid x)}{w_k \pi_{\mathrm{ref}}(y_k \mid x)} > 1 \right] \right],
\end{aligned}
\end{equation}
where $\frac{w_l}{w_k}$ is defined as:
\begin{equation}
\small
\begin{aligned}
    \frac{w_l}{w_k} = & \Bigg(\frac{\alpha(x, y_l, y_{> l})}{\alpha(x, y_k, y_{> k})} \Bigg)^{1/\beta} \cdot \prod_{i=l}^{k-1}(1-\alpha(x,y_i,y_{> i}))^{-1/\beta}.
\end{aligned}
\end{equation}

\end{theorem}
\begin{proof}

Firstly, under the PL model, we have:
\begin{equation}
\small
\mathbb{P}^*(y_i\succ\{ y_{i+1}\cdots y_M\}) = \frac{\exp(r^*(x,y_i))}{\sum_{n=i}^M \exp (r^*(x,y_n))}.
\end{equation}
Following DPO~\cite{dpo}, we can express the ground-truth reward through its corresponding optimal policy:
\begin{equation}
\small
    r^*(x,y) = \beta \log \frac{\pi^*(y|x)}{\pi_{\text{ref}}(y|x)} + \beta\log Z(x).
\end{equation}
We argue that, after thorough optimization, the optimal ranking probability $P^*(y_i\succ\{ y_{i+1}\cdots y_M\})$ derived from the optimal strategy equals the ground-truth ranking probability $\alpha(x, y_i, y_{> i})$ defined in the dataset. Then we can derive that:
\begin{equation}
\small
    \alpha(x,y_i,y_{>i}) = \frac{\exp \left( \beta \log \frac{\pi^*(y_i|x)}{\pi_{\text{ref}}(y_i|x)} \right) }{ \sum_{n=i}^M \exp \left( \beta \log \frac{\pi^*(y_n|x)}{\pi_{\text{ref}}(y_n|x)} \right)}.
\end{equation}
Rearranging, we have:
\begin{equation}
\small
\begin{aligned}
    &\frac{\alpha(x,y_l,y_{>l})}{\alpha(x,y_k,y_{>k})} \\
    = & \frac{\exp \left( \beta \log \frac{\pi^*(y_l|x)}{\pi_{\text{ref}}(y_l|x)} \right)}{\exp \left( \beta \log \frac{\pi^*(y_k|x)}{\pi_{\text{ref}}(y_k|x)} \right)} \cdot 
    \frac{\sum_{n=k}^M \exp \left( \beta \log \frac{\pi^*(y_n|x)}{\pi_{\text{ref}}(y_n|x)} \right)}{\sum_{n=l}^M \exp \left( \beta \log \frac{\pi^*(y_n|x)}{\pi_{\text{ref}}(y_n|x)} \right)} \\
    =& \frac{\exp \left( \beta \log \frac{\pi^*(y_l|x)}{\pi_{\text{ref}}(y_l|x)} \right)}{\exp \left( \beta \log \frac{\pi^*(y_k|x)}{\pi_{\text{ref}}(y_k|x)} \right)} \cdot \prod_{n=l}^{k-1} (1-\alpha(x,y_n,y_{> n})).
\end{aligned}
\end{equation}
Then we have:
\begin{equation}
\small
    \frac{\pi^*(y_l|x)}{\pi^*(y_k|x)} = \frac{w_l}{w_k} \frac{\pi_{\text{ref}}(y_l|x)}{\pi_{\text{ref}}(y_k|x)},
\end{equation}
where 
\begin{equation}
\small
\begin{aligned}
    \frac{w_l}{w_k} = & \Bigg(\frac{\alpha(x, y_l, y_{> l})}{\alpha(x, y_k, y_{> k})} \Bigg)^{1/\beta} \cdot \prod_{i=l}^{k-1}(1-\alpha(x,y_i,y_{> i}))^{-1/\beta}.
\end{aligned}
\end{equation}
If we define for each \(k=1,\ldots,M\),
\begin{equation}
\small
E_k=\Big\{ \pi^*(y_k\mid x)>\pi^*(y_j\mid x)\text{ for all } j=k+1,\ldots,K \Big\},
\end{equation}
then the top-K ranking accuracy of $\pi^*$ is given by:
\begin{equation}
\small
\mathcal{R}_{\text{KPO}}^*=P\Bigl(\bigcap_{k=1}^K E_k\Bigr).
\end{equation}

Finally, we can calculate the ranking accuracy as follows:
\begin{equation}
\small
\begin{aligned}
&\mathcal{R}_{\text{KPO}}^*(\mathcal{D}_p, \pi_{\mathrm{ref}}) 
\\& = \mathbb{E}_{ (x,y_1,\dots, y_M)\sim\mathcal{D}_p} \left[ \prod_{l=1}^{K} \prod_{k=l+1}^{M} \mathbb{I} \left[ \frac{ \pi^*(y_l \mid x)}{\pi^*(y_k \mid x)} > 1 \right] \right]
\\&= \mathbb{E}_{ (x,y_1,\dots, y_M)\sim\mathcal{D}_p} \left[ \prod_{l=1}^{K} \prod_{k=l+1}^{M} \mathbb{I} \left[ \frac{w_l \pi_{\mathrm{ref}}(y_l \mid x)}{w_k \pi_{\mathrm{ref}}(y_k \mid x)} > 1 \right] \right].
\end{aligned}
\end{equation}
This complete the proof.
\end{proof}

\subsection{Proof that KPO Outperforms S-DPO}
\label{sec:app_kpo_sdpo_proof}
Based on Theorem~\ref{th:th1}, we demonstrate in this section that KPO achieves a higher optimal ranking accuracy compared to S-DPO.

In detail, S-DPO models each data point as: $ y_1 \succ \{y_2, \dots, y_M\}$, which is a special case of KPO when $K=1$. Thus, similar to the proof of Theorem~\ref{th:th1} in Appendix~\ref{sec:proof_theorem}, we express \(\frac{\pi^*(y_l|x)}{\pi^*(y_k|x)}\) as:
\begin{equation}
\small
    \frac{\pi^*(y_l|x)}{\pi^*(y_k|x)} = \frac{w_l'}{w_k'}\frac{\pi_{\text{ref}}(y_l|x)}{\pi_{\text{ref}}(y_k|x)},
\end{equation}
where
\begin{equation}
\small
\begin{aligned}
    \frac{w_l'}{w_k'} = & \Bigg(\frac{\alpha(x, y_l, y_{\geq l})}{\alpha(x, y_k, y_{> k})} \Bigg)^{1/\beta} \cdot \prod_{i=l}^{k-1}(1-\alpha(x,y_i,y_{> i}))^{-1/\beta}\\ 
    & \cdot \mathbb{I}[l=1] + \mathbb{I}[l\neq 1].
\end{aligned}
\end{equation}

As a result, the optimal ranking accuracy of S-DPO is:
\begin{equation}
\small
\begin{aligned}
&\mathcal{R}_{\text{S-DPO}}^*(\mathcal{D}_p, \pi_{\mathrm{ref}}) 
\\&= \mathbb{E}_{ (x,y_1,\dots, y_M)\sim\mathcal{D}_p} \left[ \prod_{l=1}^{K} \prod_{k=l+1}^{M} \mathbb{I} \left[ \frac{w_l' \pi_{\mathrm{ref}}(y_l \mid x)}{w_k' \pi_{\mathrm{ref}}(y_k \mid x)} > 1 \right] \right].
\end{aligned}
\end{equation}

Next, we aim to prove that \(\frac{w_l}{w_k} > \frac{w_l'}{w_k'}\) for all \( l \in \{2, \dots, K\} \) and \( k \in \{l + 1, \dots, M\} \).

Since the ranking probabilities $\alpha(x,y_i,y_{>i})$ are provided by the dataset $\mathcal{D}_p$, this implies that
\begin{equation}
\small
    r^*(x,y_l)>r^*(x,y_k), \forall l<k.
\end{equation}
Hence, we can derive that:
\begin{equation} 
\small
\Bigg(\frac{\alpha(x, y_l, y_{> l})}{\alpha(x, y_k, y_{> k})} \Bigg)^{1/\beta} \cdot \prod_{i=l}^{k-1}(1-\alpha(x,y_i,y_{> i}))^{-1/\beta} > 1.
\end{equation}

Therefore, we conclude that \(\frac{w_l}{w_k} > \frac{w_l'}{w_k'}\) for all \( l \in \{2, \dots, K\} \) and \( k \in \{l + 1, \dots, M\} \).

Subsequently, for $l\neq 1$, we have:
\begin{equation}
\small
\begin{aligned}
 \mathbb{I} \left[ \frac{w_l \pi_{\mathrm{ref}}(y_l \mid x)}{w_k \pi_{\mathrm{ref}}(y_k \mid x)} > 1 \right] > \mathbb{I} \left[ \frac{w_l' \pi_{\mathrm{ref}}(y_l \mid x)}{w_k' \pi_{\mathrm{ref}}(y_k \mid x)} > 1 \right].
\end{aligned}
\end{equation}

Therefore, we conclude that $\mathcal{R}_{\text{KPO}}^*(\mathcal{D}_p, \pi_{\text{ref}}) > \mathcal{R}_{\text{S-DPO}}^*(\mathcal{D}_p, \pi_{\text{ref}})$.

\subsection{Optimization Objective of KTO}
\label{sec:kto}
In this section, we provide a detailed introduction to the optimization objectives of KTO~\cite{kto}.

Given that $\lambda_D$ and $\lambda_U$ are hyperparameters for desirable and undesirable outputs respectively, the KTO loss is defined as:
\begin{equation}
\small
   \mathcal{L}_{\rm KTO}(\pi_\theta;\pi_{\rm ref}) = 
     \mathbb{E}_{x,y \sim \mathcal{D}} [ \lambda_y - v_{\text{KTO}}(x, y) ],
    \label{eq:kto_loss}
\end{equation}
where 
\begin{equation*}
\small
\begin{split}
    r_\theta(x, y) &= \log \frac{\pi_\theta(y|x)}{\pi_\text{ref}(y|x)} \\
    z_0 &= \text{KL}(\pi_{\theta}(y'|x)\|\pi_\text{ref}(y'|x)) \\
    v_{\text{KTO}}(x, y) &=
    \begin{cases}
    \lambda_D \sigma(\beta(r_\theta(x,y) - z_0)) \ \text{if } y \sim y_\text{desirable}|x \\
    \lambda_U \sigma(\beta(z_0 - r_\theta(x,y))) \ \text{if } y \sim y_\text{undesirable}|x\\
    \end{cases}
\end{split}
\end{equation*}

\section{Ground Truth Label}
\label{sec:ground_truth_label}
As mentioned in Section~\S\ref{sec:query_adaptive_k}, we need to use the ground truth labels in the dataset to re-rank the top-$K$ items.
In practice, ground truth relevance labels are derived as follows:  
\begin{itemize}[topsep=0pt,itemsep=0pt,parsep=0pt,leftmargin=*]
\item \textit{Product Search Tasks}: Each candidate item is assigned a relevance score with respect to the query, typically from a discrete set such as $\{0, 1, 2, 3\}$. 
\item \textit{Recommendation Tasks}: Only the item most recently interacted with by the user is typically considered relevant, while all other items are treated as irrelevant. This scenario can be seen as a special case where one item's relevance score is ``1'', and all others are assigned a score of ``0''.
\end{itemize}

\section{Experimental Settings}
\subsection{Datasets}
\label{sec:app_datasets}
In this section, we provide a detailed description of three datasets, as outlined below. The statistical information is presented in Table~\ref{tab:dataset_statistics}.
\begin{itemize}[topsep=0pt,itemsep=0pt,parsep=0pt,leftmargin=*]
\item \textit{MovieLens}: This is a widely used dataset for movie recommendation tasks, containing user ratings for various movies and offering subsets of different sizes. Given the substantial computational demands of LLMs, we chose the MovieLens100K dataset for our experiments.
\item \textit{Goodreads}: This dataset comprises user ratings and reviews of books. To manage the dataset size, we filtered out users with fewer than 20 interactions on Goodreads.
\item \textit{Shopping Queries}: This dataset features a collection of challenging Amazon search queries and corresponding results. To limit its size, we excluded products associated with fewer than 5 queries.
\end{itemize}
\begin{table}[htbp]
\centering

 \renewcommand\arraystretch{0.9} 
 \small
\begin{tabular}{lrrr}
\toprule
Dataset &\#Query &\#Item &\#Interaction             \\ 
\toprule
MovieLens&943&1,682&100,000 \\
Goodreads&6,031&4,500&220,100 \\
Shopping Queries&21,852&12,882&96,788 \\
\bottomrule
\end{tabular}
\caption{\textbf{Statistics of datasets.}}
\label{tab:dataset_statistics}
\end{table}

\subsection{Baselines}
\subsubsection{Preference Alignment Methods}
\label{sec:app_preference_baseline}
We compare KPO with various preference alignment methods, including KTO~\cite{kto}, DPO~\cite{dpo}, SimPO~\cite{simpo}, Conservative DPO (cDPO)~\cite{cdpo}, S-DPO~\cite{softmaxdpo}, and DPO$_{\text{PL}}$~\cite{dpo}. Detailed descriptions of these methods are provided below:
\begin{itemize}[topsep=0pt,itemsep=0pt,parsep=0pt,leftmargin=*]
\item \textit{KTO}: Inspired by Kahneman and Tversky's prospect theory~\cite{kt1,kt2}, this method relies solely on binary labels, classifying samples as either "good" or "bad," which can be considered a point-wise approach.
\item \textit{DPO}: Provides a closed-form solution for the reward model in RLHF~\cite{rlhf} and enables offline optimization of the pair-wise preference model.
\item \textit{SimPO}: Proposes a simplified optimization algorithm compared to DPO, eliminating the need for a reference model.
\item \textit{Conservative DPO (cDPO)}: Introduces a hyperparameter $\epsilon$ to account for the flip rate of noisy labels.
\item \textit{S-DPO}: Incorporates multiple negative samples in user preference data and develops an alternative DPO loss formulation tailored for LM-based recommenders, linked to softmax sampling strategies.
\item \textit{DPO$_{\text{PL}}$}: Extends DPO's Bradley-Terry modeling to the list-wise Plackett-Luce modeling.
\end{itemize}

\subsubsection{Recommendation Models}
\label{sec:app_rec_models}
\begin{table*}[htbp]

\centering
\small
\begin{tabular}{l|ccccc|ccccc}
\toprule
\multirow{2}{*}{Method} & \multicolumn{5}{c|}{MovieLens} &\multicolumn{5}{c}{Goodreads}\\
 &HR@1 & HR@5 & HR@10 & N@5 & N@10 &HR@1 & HR@5 & HR@10 & N@5 & N@10\\
 \midrule
SFT & 0.5053& 0.8526& 0.9368& 0.6983& 0.7255& 0.4809& 0.8369& 0.9468& 0.6675&0.7034\\
Random & 0.5579& 0.8842& \textbf{0.9684}& 0.7361& 0.7620& 0.5042& 0.8719& 0.9584& 0.6994& 0.7272\\
Descending & 0.5474& 0.8737& \textbf{0.9684}& 0.7233& 0.7532& 0.4942& 0.8686& 0.9584& 0.6949& 0.7239\\
Ascending & \textbf{0.5684}& \textbf{0.8947}& \textbf{0.9684}& \textbf{0.7381}& \textbf{0.7637}& \textbf{0.5158}& \textbf{0.8735}& \textbf{0.9667}& \textbf{0.7024}&\textbf{0.7353}\\
\bottomrule
\end{tabular}
\caption{\textbf{Comparison of different training data orders.} Bold indicates the best performance.}
\label{tab:curriculum_learning}
\end{table*}
We compare KPO with various recommendation models, which can be broadly classified into two categories: traditional models and LLM-based models.

The traditional recommendation models include:
\begin{itemize}[topsep=0pt,itemsep=0pt,parsep=0pt,leftmargin=*]
\item \textit{SASRec}: An attention-based sequential recommendation model designed to effectively capture long-range semantic dependencies in user behavior sequences.
\item \textit{GRU4Rec}: A recurrent neural network (RNN)-based model known for its simplicity and efficiency in recommendation tasks.
\item \textit{Caser}: A convolutional neural network (CNN)-based model that interprets a user's historical behavior sequence as an ``image'' and leverages CNN operations to extract meaningful patterns.
\end{itemize}

The LLM-based recommendation models include:
\begin{itemize}[topsep=0pt,itemsep=0pt,parsep=0pt,leftmargin=*]
\item \textit{MoRec}: A model that enhances traditional recommendation models by integrating modality-specific features of items.
\item \textit{LLaRA}: A hybrid model that combines LLM with the traditional models' embeddings through hybrid item representations.
\end{itemize}

\subsubsection{LLM-based Rankers}
\label{sec:app_rank_models}
We compare KPO with various LLM-based rankers, including RankGPT$_\text{3.5}$~\cite{rankgpt}, and LlamaRec~\cite{llamarec}.
Detailed descriptions of these rankers are provided below:
\begin{itemize}[topsep=0pt,itemsep=0pt,parsep=0pt,leftmargin=*]
\item \textit{RankGPT$_\text{3.5}$}: RankGPT directly prompts ChatGPT~\cite{chatgpt} to rank a list of candidate items in a zero-shot manner.
\item \textit{LlamaRec}: LlamaRec ranks candidate items based on the logits output by the model.
\end{itemize}

\subsection{Implementation Details}
\label{sec:app_implemetation}
Our experiments are conducted on eight NVIDIA A40 GPUs. For the KPO method, we use the LLama-3.2-3B-Instruct~\cite{llama32} model as the backbone and apply LoRA~\cite{lora} for fine-tuning. Specifically, the LoRA rank is set to 32, and the LoRA alpha is configured to 64. 
During the supervised fine-tuning (SFT) stage, the model is trained for 5 epochs with a learning rate of 1e-4. In the preference alignment stage, the learning rate is reduced to 1e-5, and training is performed over 3 epochs. The global batch size is fixed at 128. To ensure optimal performance, we select the model checkpoint that achieves the best results on the validation set. 
Additionally, a warm-up strategy is employed, where the learning rate is initialized to $\frac{1}{100}$ of its maximum value and gradually increased using a cosine scheduler.
For traditional models, we adopt the settings outlined in~\cite{frame_seq_rec}, using a learning rate of 0.001, an embedding dimension of 64, and a batch size of 256. To determine the optimal L2 regularization coefficient, we conduct a grid search over the values $[1e-3, 1e-4, 1e-5, 1e-6, 1e-7]$.
For other LLM-based models, we follow the training protocol described in LLaRA~\cite{llara}, training the models for up to 5 epochs with a batch size of 128.

\section{Additional Experiments}
\subsection{Analysis of Time Complexity}
\label{sec:app_time_cost}
As mentioned in Section~\S\ref{sec:method_time_cost}, the additional $K$-layer loop introduced by KPO, compared to S-DPO, does not significantly increase the actual runtime. To support this claim, we conducted experiments to compare the runtime performance of KPO and S-DPO in practice.

We measured the average runtime of each phase during optimization on the MovieLens dataset using an NVIDIA A40 GPU with a batch size of 4. As shown in Table~\ref{tab:time_cost}, KPO's total runtime is only 2\% longer than S-DPO. This slight increase arises from Phases 1 and 3 dominating the computation, while the added complexity in Phase 2 has minimal impact. Therefore, KPO achieves runtime efficiency comparable to S-DPO despite its higher theoretical complexity.
\begin{table}[htbp]
\centering
\small

 \renewcommand\tabcolsep{1pt} 
\begin{tabular}{l|cc|cccc}
\toprule
\multirow{2}{*}{Method}& \multicolumn{2}{c|}{Complexity}& \multicolumn{4}{c}{Runtime} \\
& Phase 1 & Phase 2  & Phase 1 & Phase 2 & Phase 3&Total \\
\toprule
S-DPO & $\Theta(M)$ & $\Theta(M)$  &2.82 & 0.03& 8.04& 10.89\\
KPO & $\Theta(M)$ & $\Theta(K\cdot M)$  &2.82 & 0.24& 8.04& 11.10\\
\bottomrule
\end{tabular}
\caption{\textbf{Results of time complexity and actual runtime.  
} ``Complexity'' refers to the number of iterations per phase, with execution time measured in seconds.}
\label{tab:time_cost}
\end{table}
\begin{table}[htbp]

\centering
 \renewcommand\tabcolsep{2.4pt} 
 \renewcommand\arraystretch{1.2} 
\begin{tabular}{l|ccccc}
\toprule
\multirow{2}{*}{Method} &\multicolumn{5}{c}{Shopping Queries}\\
 &HR@1 & HR@5 & HR@10 & N@5 & N@10  \\
\midrule
KTO &  0.5120& 0.8430& 0.9510& 0.6891& 0.7243\\
DPO &  0.5210& 0.8560& 0.9550& 0.6968& 0.7288\\
SimPO &  0.5210& 0.8580& 0.9630& 0.6991& 0.7331\\
cDPO &  0.5240& 0.8520& 0.9540& 0.6972& 0.7304\\
S-DPO &  0.5270& 0.8420& 0.9510& 0.6936& 0.7293\\
DPO$_{\text{PL}}$ & 0.5230& 0.8560& 0.9530& 0.6997& 0.7315\\
KPO$_{\text{CUT}}$ & 0.5220& 0.8410& 0.9480& 0.6929& 0.7279\\
\midrule
KPO & \textbf{0.5330}& \textbf{0.8670}& \textbf{0.9670}& \textbf{0.7087}& \textbf{0.7414}\\
\bottomrule
\end{tabular}
\caption{\textbf{Comparison for optimization objectives on the Shopping Queries dataset.} Bold indicates the best performance.}
\label{tab:esci_single_compare}
\end{table}
\subsection{Shopping Queries Dataset with One Ground Truth Item}
\label{sec:esci_single}
We also align the experimental setup of the Shopping Queries dataset with that of the recommendation dataset: a candidate item list is composed of one ground truth item and 19 randomly sampled items.
The experimental results are presented in Table~\ref{tab:esci_single_compare}.

Based on the experimental results, we can conclude that KPO outperforms other preference alignment methods, which demonstrates the effectiveness of KPO.

\subsection{$K$-aware Curriculum Learning}
\label{sec:app_cl}
To demonstrate the effectiveness of $K$-aware curriculum learning, we present the performance of three training data orders—random, descending, and ascending—on the MovieLens and Goodreads test set.
For better comparison, we also present the performance of the SFT model.  
The experimental results are summarized in Table~\ref{tab:curriculum_learning}. 

From these results, we have drawn the following findings and conclusions:  
The model trained on ``Ascending'' data consistently outperforms those trained on ``Random'' and ``Descending'' data.
This indicates that starting with simpler data and gradually progressing to more complex data is beneficial for improving model performance.

\end{document}